\documentclass[pre,amsmath,twocolumn]{revtex4}
\pdfoutput=1
\usepackage{bm}
\usepackage{amssymb}
\usepackage{graphicx}
\usepackage{subfigure}
\usepackage{pdfpages}
\def \bea{\begin{eqnarray}}
\def \eea{\end{eqnarray}}

\begin{document}
\renewcommand{\thefootnote}{\Roman{footnote}}

\title{Jamming vs. Caging in Three Dimensional Jamming Percolation}

\author{Nimrod Segall$^{1,2}$}
\email{nimrods6@mail.tau.ac.il}
\author{Eial Teomy$^{2,3}$}
\email{eialteom@tau.ac.il}
\author{Yair Shokef$^{2,3}$}
\email{shokef@tau.ac.il}
\affiliation{$^1$School of Chemistry, $^2$Sackler Center for Computational Molecular and Materials Science, and $^3$School of Mechanical Engineering, Tel Aviv University, Tel Aviv 6997801, Israel}

\begin{abstract}

We investigate a three-dimensional kinetically-constrained model that exhibits two types of phase transitions at different densities. At the jamming density $\rho_J$ there is a mixed-order phase transition in which a finite fraction of the particles become frozen, but the other particles may still diffuse throughout the system. At the caging density $\rho_C>\rho_J$, the mobile particles are trapped in finite cages and no longer diffuse. The caging transition occurs due to a percolation transition of the unfrozen sites, and we numerically find that it is a continuous transition with the same critical exponents as random percolation. 

\end{abstract}

\maketitle

\section{Jamming percolation}

Kinetically-constrained models are widely used in the study of jamming- and glass-transitions in amorphous materials~\cite{FA,KA,Jackle_Kronig,Kronig_Jackle,KCM_rev1,KCM_rev2}. These lattice-gas (or spin-$\frac{1}{2}$ Ising) models are characterized by having some local restricting set of rules for particle movement (or spin flip). In spin-facilitated models, the two possible states at each site represent active versus inactive regions, with the fraction of active sites increasing with increasing temperature. The density of these active sites corresponds to the density of vacancies in lattice-gas models. In the Fredrickson-Andresen~\cite{FA} and Kob-Andersen~\cite{KA} models for example, the number of neighboring vacant (active) sites sets whether a site is \emph{blocked} or not, and \emph{jamming} of the system is often studied in terms of the fraction of \emph{frozen} particles that will permanently remain blocked under the model's dynamics. In the thermodynamic limit these models jam only in the pathological limit of zero temperature or alternatively vanishing vacancy density~\cite{FA,Holroyd_2003,TBF_framing,Balogh}, thus only spatial confinement may induce jamming at a nontrivial density in them~\cite{Teomy_PRE2012,Teomy_PRE2014}. 

\begin{figure}[b]
\includegraphics[width=\columnwidth]{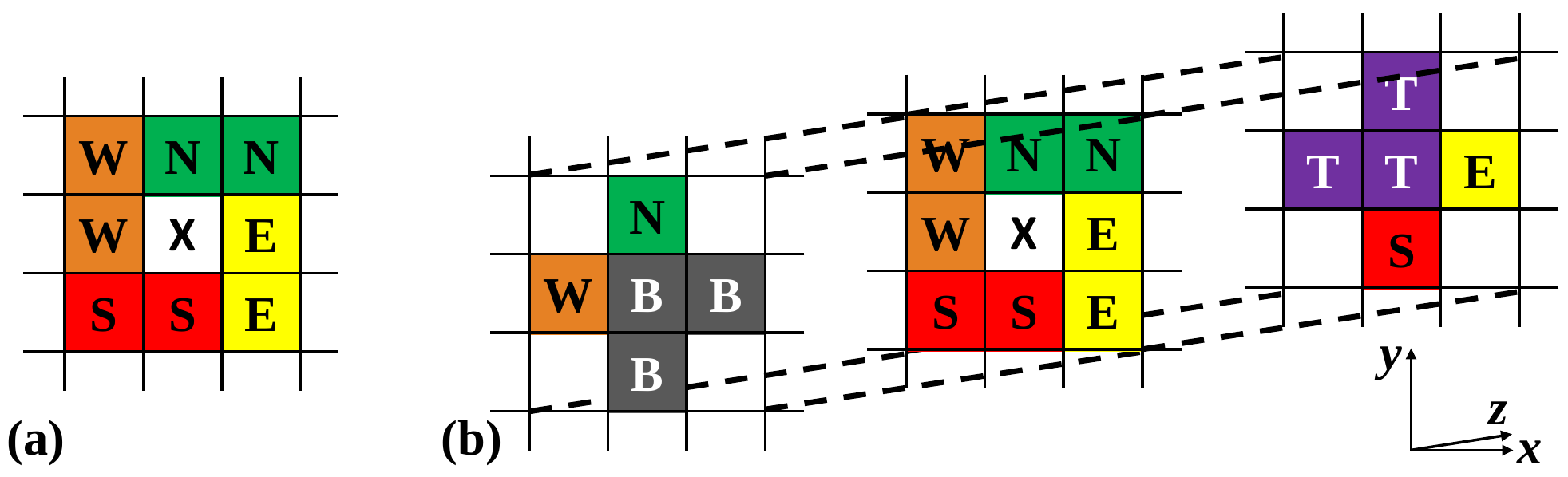}
\caption{\textbf{Kinetic rules:} (a) In the 2D spiral model a site ($\times$) is unblocked if its ((W {\it or} E) {\it and} (S {\it or} N)) sets are completely empty. (b) In our 3D model the ($\times$) site is unblocked if the sets ((W {\it or} E) {\it and} (S {\it or} N) {\it and} (B {\it or} T)) are completely empty.}
\label{fig:spiral_rules}
\end{figure}

In a new class of kinetically-constrained models, referred to as \emph{jamming-percolation models}~\cite{Knights,JS_Comment,TBF_Reply,Schwarz_force_balance}, the kinetic rules depend not only on the number of neighboring active (or vacant) sites but also on their relative orientations. These models are constructed such that already at some nontrivial density of inactive (or occupied) sites $\rho_J<1$, there are percolating clusters of permanently-frozen particles, and the system becomes jammed in the sense that in the thermodynamic limit there is a finite fraction of permanently frozen particles that will never move (or flip). We focus on a three-dimensional (3D) extension~\cite{Ghosh} of the two-dimensional (2D) spiral model~\cite{TBF_Reply,Spiral_EPJB, Spiral_JSP,Shokef_Liu,Cugliandolo_Corberi}. The spiral model is defined on a square lattice by having a set of kinetic constraints so that if (\emph{N} or \emph{S}) and (\emph{W} or \emph{E}) sets are completely empty the central site is unblocked, see Fig.~\ref{fig:spiral_rules}. The 3D model is defined on a cubic lattice so that if (\emph{N} or \emph{S}) and (\emph{W} or \emph{E}) and (\emph{T} or \emph{B}) sets are completely empty the central site is unblocked~\cite{proof_wrong}.

In the spin versions of the models, unblocked spins may stochastically flip between their active and inactive states at some temperature-dependent rates. In the lattice-gas versions, an unblocked particle may hop to a neighboring vacant site only if that particle is unblocked by the model rules also after moving to that target site. These kinetic rules are constructed so that the dynamics will obey time-reversal symmetry, and as a consequence a particle can change with time from being blocked to being unblocked, however it cannot change with time between being frozen and unfrozen. Thus the initial condition sets which particles are frozen and which are unfrozen, or \emph{mobile}. The behavior in the spin-facilitated and in the lattice-gas versions of these models is almost identical, and we will focus on the lattice-gas models, for which particles move on the lattice. Here we can study whether the mobile particles exhibit long-time diffusive behavior or whether they are \emph{caged} by the frozen particles. The 2D and the 3D models undergo a mixed-order phase transition at some (different) jamming density $\rho_J$, at which the fraction of frozen particles jumps discontinuously from zero to some finite value~\cite{Spiral_EPJB,Ghosh}, as in a first-order transition, while the minimal number of moves required to unblock a site diverges~\cite{Shokef_Liu,Ghosh}, reflecting diverging length and time scales~\cite{Teomy_PRE2015} as in a second-order transition. 

\section{Caging}

\begin{figure}[t]
\includegraphics[width=\columnwidth]{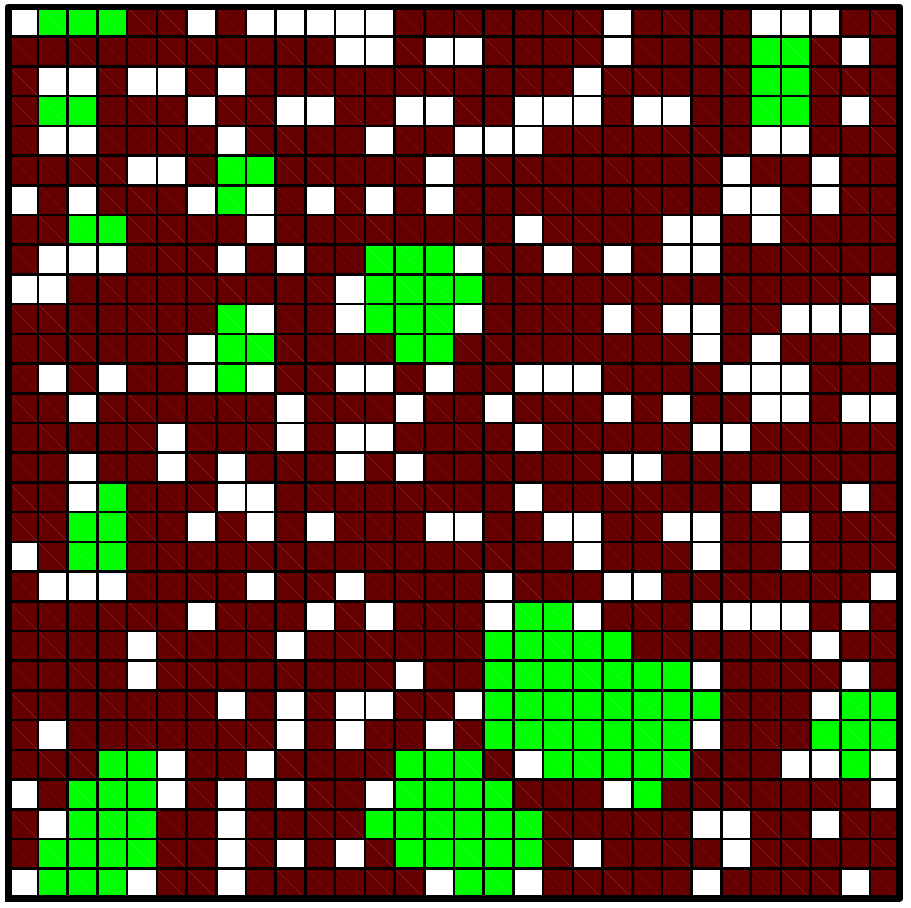}
\caption{\textbf{Jamming is caging in 2D:} A representative configuration of the 2D spiral model at $\rho = 0.69$ for a square lattice of linear size $L=30$. Frozen particles are marked in dark red, and unfrozen sites in bright green. At the critical jamming density $\rho_J \approx 0.68$, frozen particles appear and immediately span the system and cage the unfrozen particles.}
\label{fig:cluster_map_2d}
\end{figure}

Interestingly, in these lattice-gas models the motion of the unfrozen particles exhibits a qualitative difference between two and three dimensions; since the kinetic rules in jamming-percolation models map to a \emph{directed-percolation} problem, the percolating clusters of \emph{frozen sites} have a string-like, or quasi-one-dimensional (1D) structure. In two dimensions, once frozen particles appear, their 1D strings \emph{cage} the unfrozen particles into compact regions and thus the self diffusion vanishes at $\rho_J$ and jamming and caging are tantamount, see Fig.~\ref{fig:cluster_map_2d}. The frozen clusters are 1D also in three dimensions, however here the unfrozen regions may be topologically connected and can allow unfrozen particles to use the third dimension in order to diffuse with time, and indeed dynamical simulations demonstrated long-time diffusive behavior also in some range of $\rho>\rho_J$~\cite{Ghosh}.

In this paper we show that in 3D jamming percolation the self-diffusion coefficient vanishes at some second, higher critical density $\rho_J<\rho_C<1$, and that this second phase transition is a continuous percolation transition. We first identify not only the frozen particles but also the frozen vacant sites, namely sites that are empty and no particle will ever be able to enter them. Together, the frozen particles and the frozen vacant sites constitute the entirety of frozen sites in the system. Secondly, we infer on the existence of long-time diffusive motion of the unfrozen particles based on whether the unfrozen sites percolate across the system. We find that in two dimensions caging coincides with the jamming transition, while in three dimensions caging occurs only at a second critical density $\rho_C$. Namely, \emph{there is a range of densities $\rho_J < \rho < \rho_C$ in which the system is jammed yet uncaged}.

\begin{figure}[t]
\includegraphics[width=0.9\columnwidth]{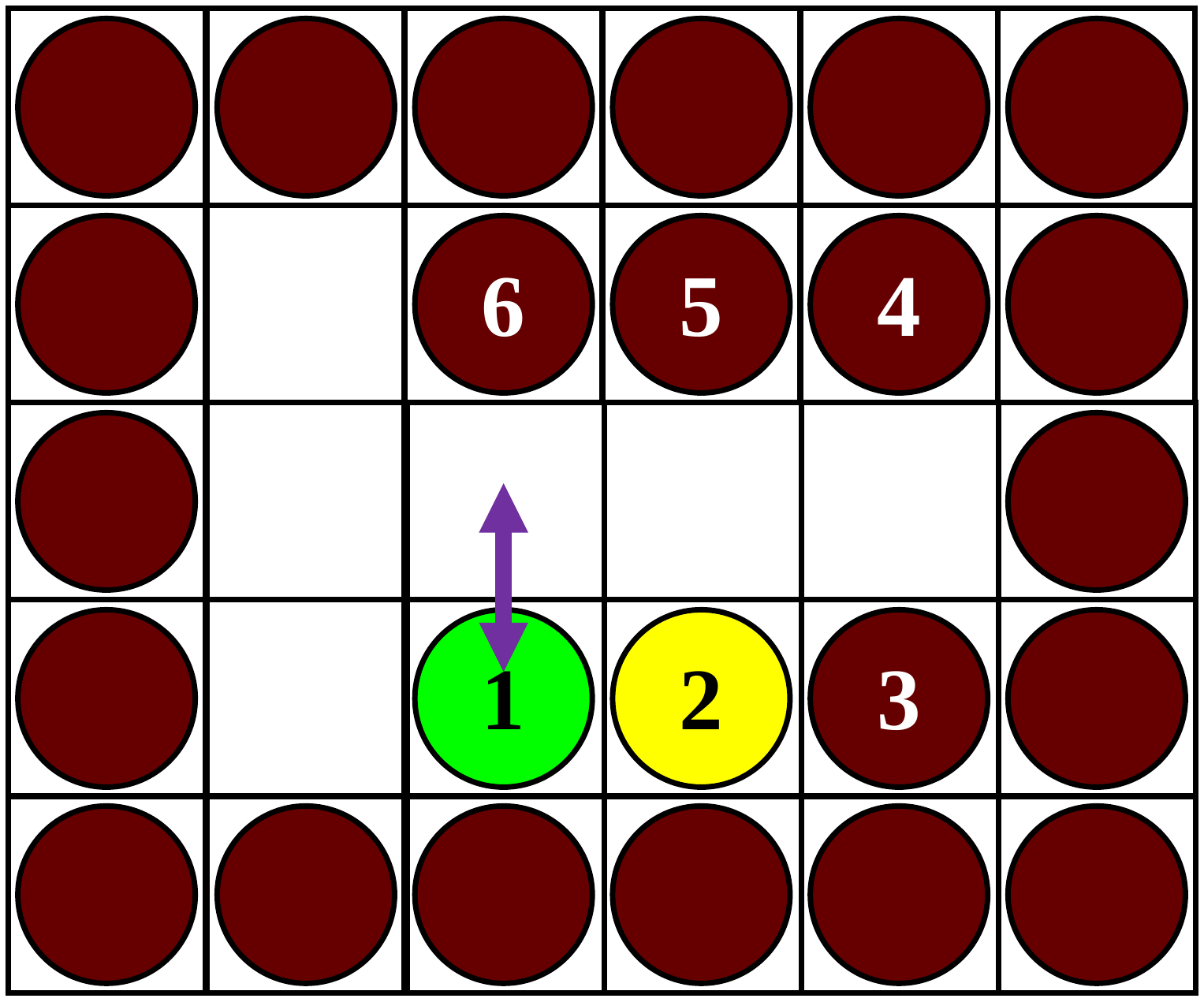}
\caption{\textbf{Culling vs. dynamics:} Configuration for which culling fails to exactly identify the frozen particles in the lattice-gas spiral model. In this small region surrounded by occupied sites (solid circles) the only possible dynamics are particle 1 moving one site up and then back down, thus it is the only unfrozen particle. However, after particle 1 is culled, particle 2 is culled and is identified erroneously as unfrozen. Particles 3-6 are identified as frozen both by the culling algorithm and by analyzing the actual dynamics.}
\label{fig:culling_wrong}
\end{figure}

\section{Frozen particles and unfrozen sites}\label{sec:unfrozen_sites}

We begin with a \emph{culling} procedure to identify the frozen particles. In this process a random initial configuration on the lattice is considered, we then test according to the kinetic constraints of the model's dynamics whether each particle is blocked or not, and subsequently remove the unblocked particles. The resulting configuration is then analyzed again in the same manner to identify the blocked particles, and the process continues iteratively until no more particles may be removed. We identify the remaining particles as the frozen particles that will never be able to move. For spin-facilitated models, such culling generates one specific possible trajectory of the dynamics and thus exactly identifies the permanently-frozen inactive spins. For lattice-gas models on the other hand, culling formally gives only a lower bound on the frozen particles, and in principle particles that were culled may in fact be frozen, see Fig.~\ref{fig:culling_wrong}. However, in the thermodynamic limit we expect such effects to be negligible. Moreover, we speculate that the culling algorithm might misidentify frozen particles as unfrozen only if they are caged. This misidentification occurs when a particle is mobile but localized, in the manner that no matter where it goes it still blocks one other particle, as shown in Fig.~\ref{fig:culling_wrong}. If the mobile particle is not caged, then it can move far from its initial position, and thus it no longer blocks the other particle. Since we are interested in the motion of particles in the uncaged case, this discrepancy should not affect our results.

\begin{figure}[t]
\includegraphics[width=0.75\columnwidth]{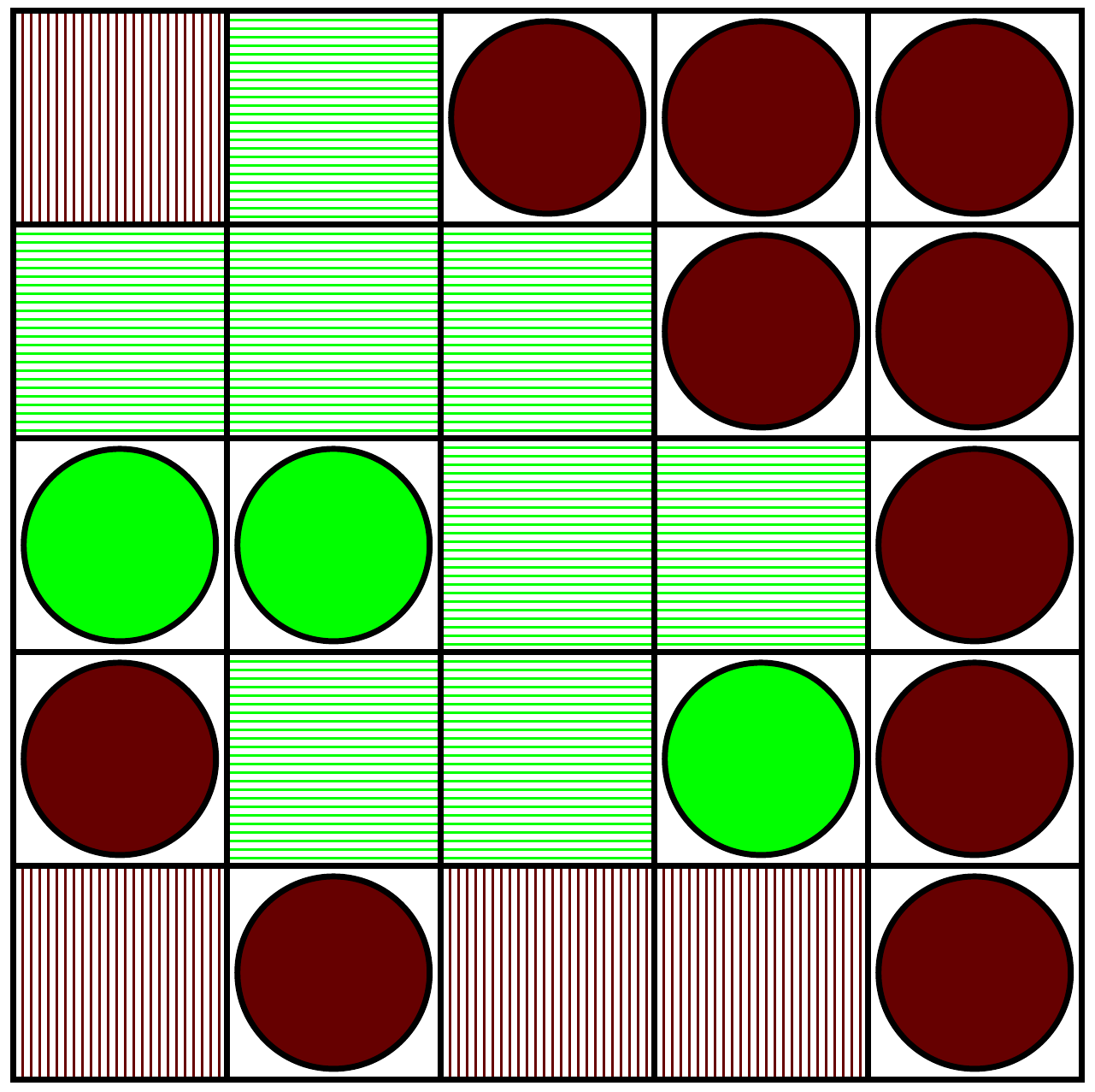}
\caption{\textbf{Frozen particles and sites:} We first run culling on a configuration, to identify the frozen particles (dark red circles) and the unfrozen particles (bright green circles). Then, we identify the frozen sites (red vertical shading) and the unfrozen sites (green horizontal shading). In this example, analysis was performed using the kinetic rules of the spiral model, and assuming periodic boundary conditions.}
\label{fig:culling_explain}
\end{figure}

Culling is very useful as it allows one to infer on the very long-time behavior of a slowly-relaxing system without running a dynamical simulation, and by only analyzing the structure of its initial configuration~\cite{Teomy_PRE2015}. Similarly, \emph{we would like to predict the long-time diffusive behavior of mobile, or unfrozen particles in our system based solely on the structure of the initial configuration}. This would be very valuable since measuring the self-diffusion of particles is time consuming and indecisive. The first step toward this goal is identifying the vacant frozen sites, namely sites that will never be occupied by a particle. These are vacant sites that will never be able to change their occupation and become occupied because of the presence of frozen particles in their proximity. Therefore, we consider the frozen particles as identified by the culling algorithm and check for all the initially-vacant sites whether the frozen particles block them or not, see Fig.~\ref{fig:culling_explain}. The resulting unfrozen vacant sites together with the initial locations of the unfrozen particles constitute the unfrozen sites in the system, in which the mobile particles can move.

\begin{figure}[t]
\includegraphics[width=0.397\columnwidth]{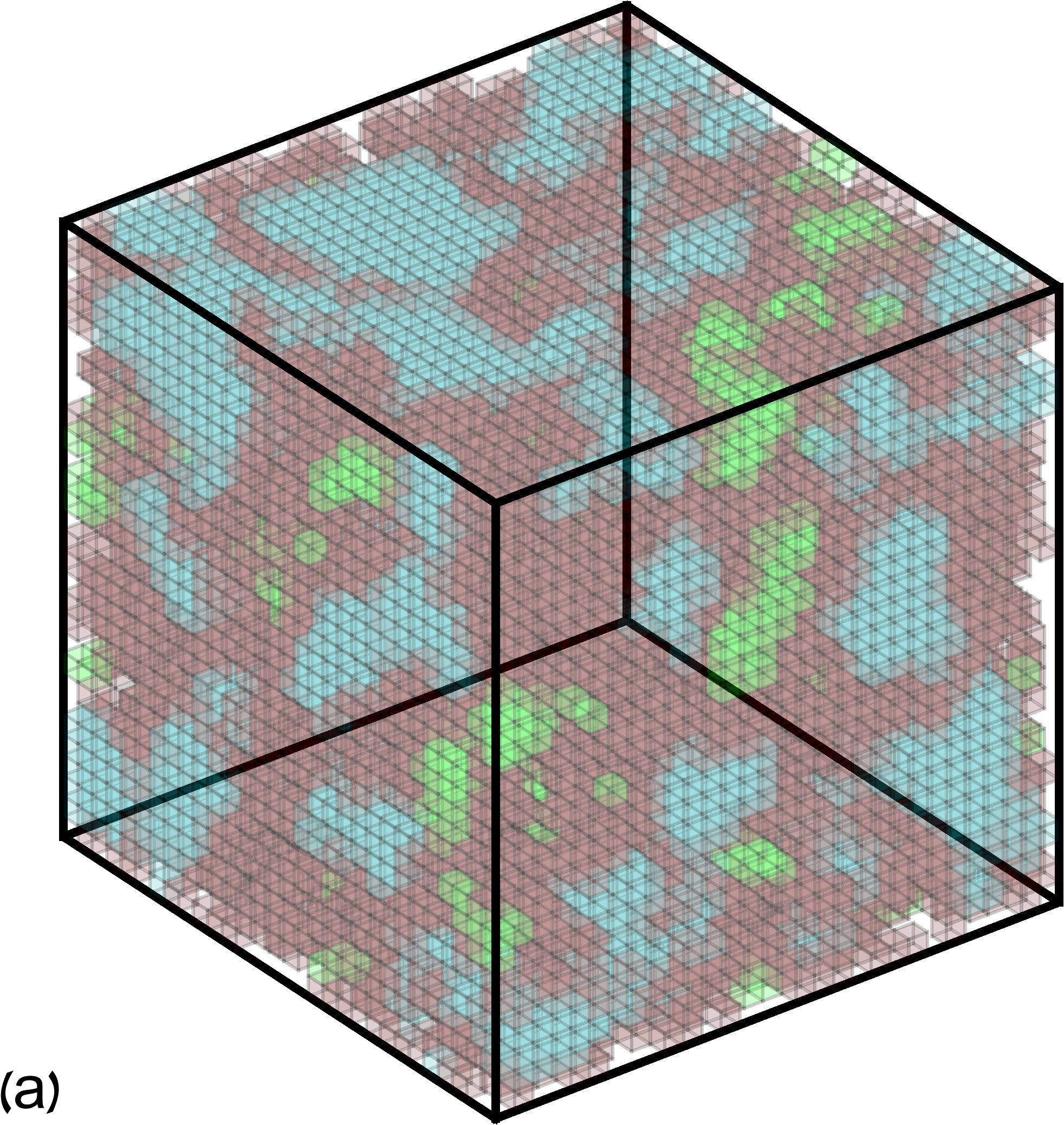}
\includegraphics[width=0.397\columnwidth]{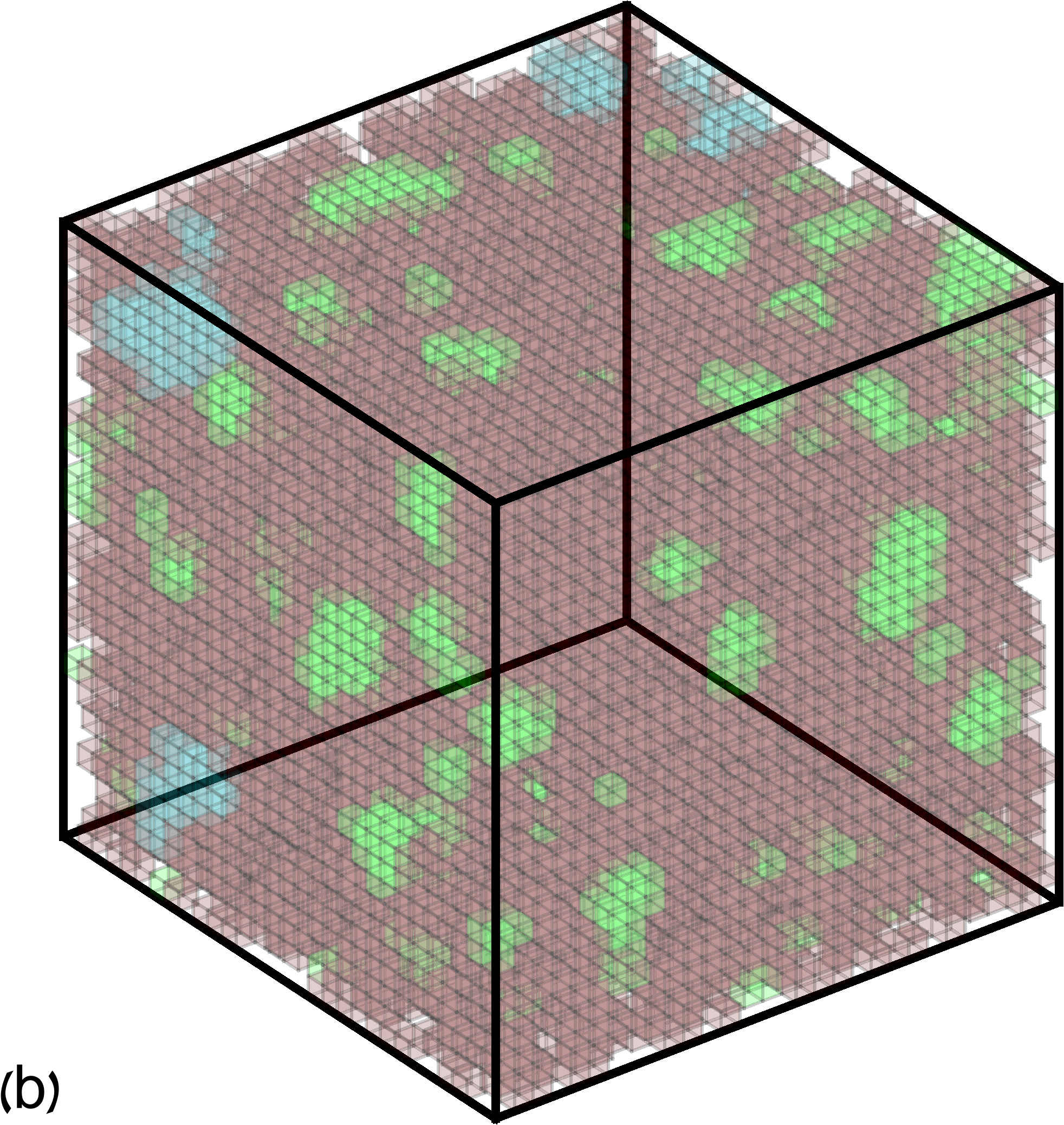}
\includegraphics[width=0.397\columnwidth]{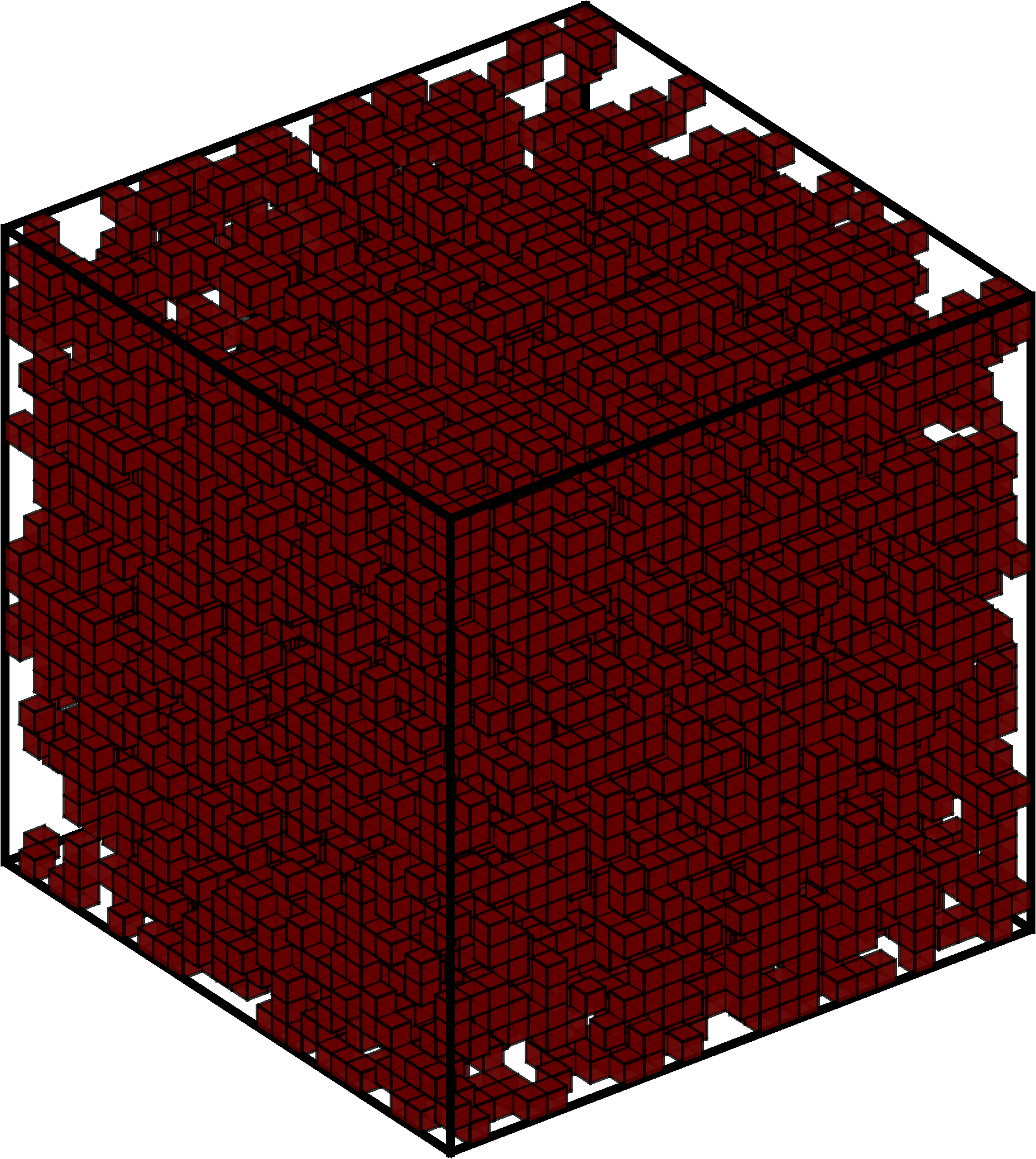}
\includegraphics[width=0.397\columnwidth]{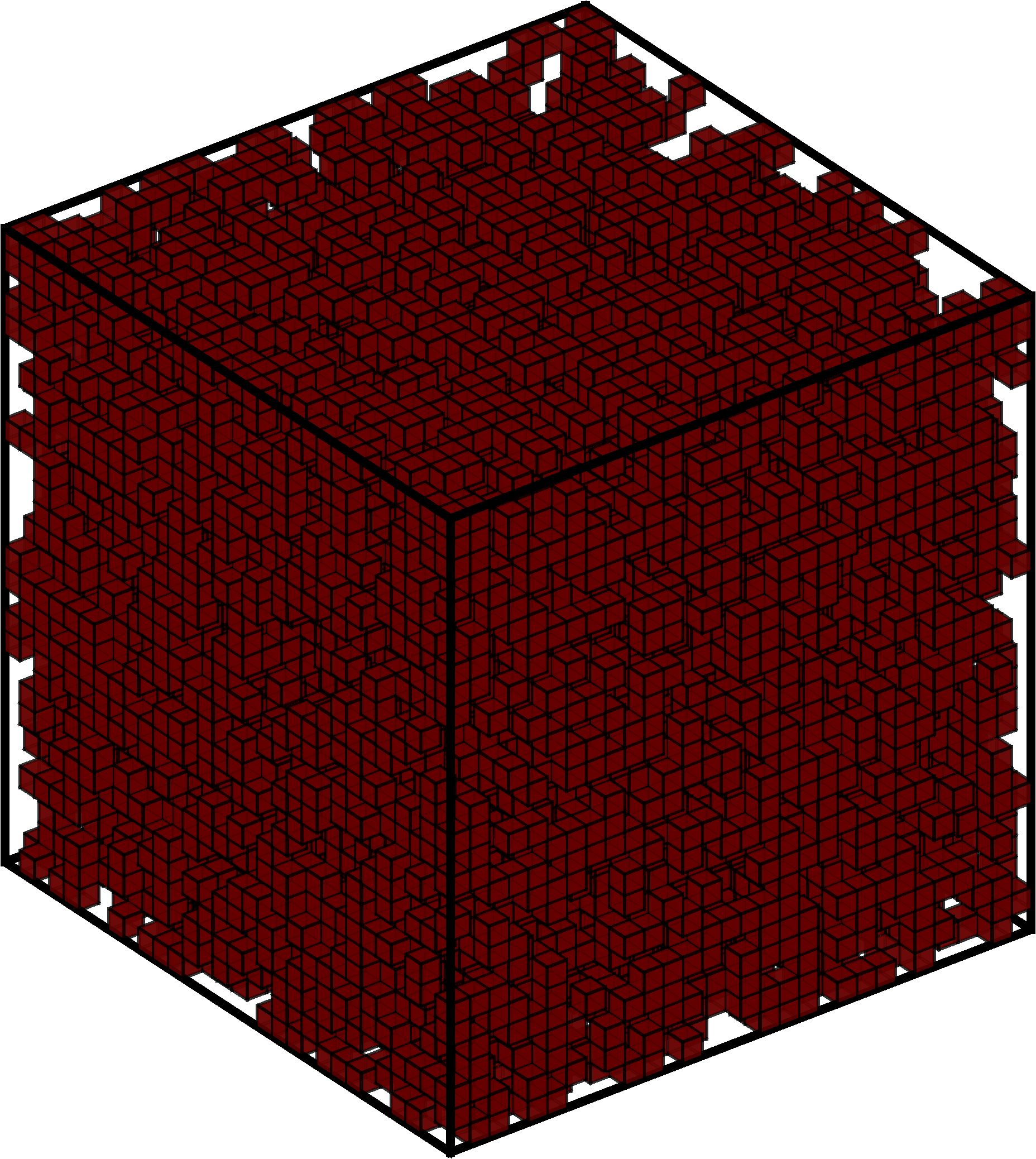}
\includegraphics[width=0.397\columnwidth]{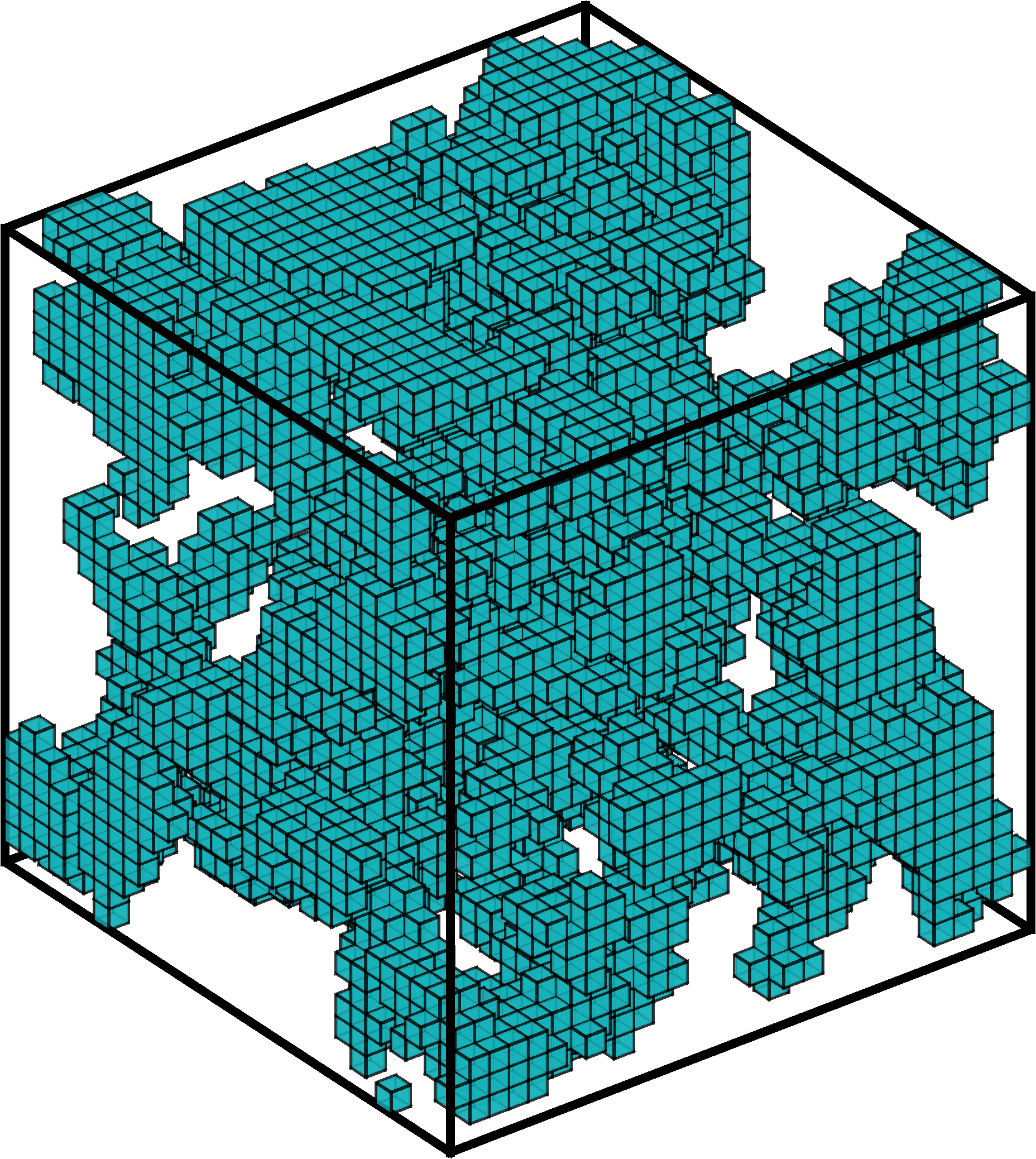}
\includegraphics[width=0.397\columnwidth]{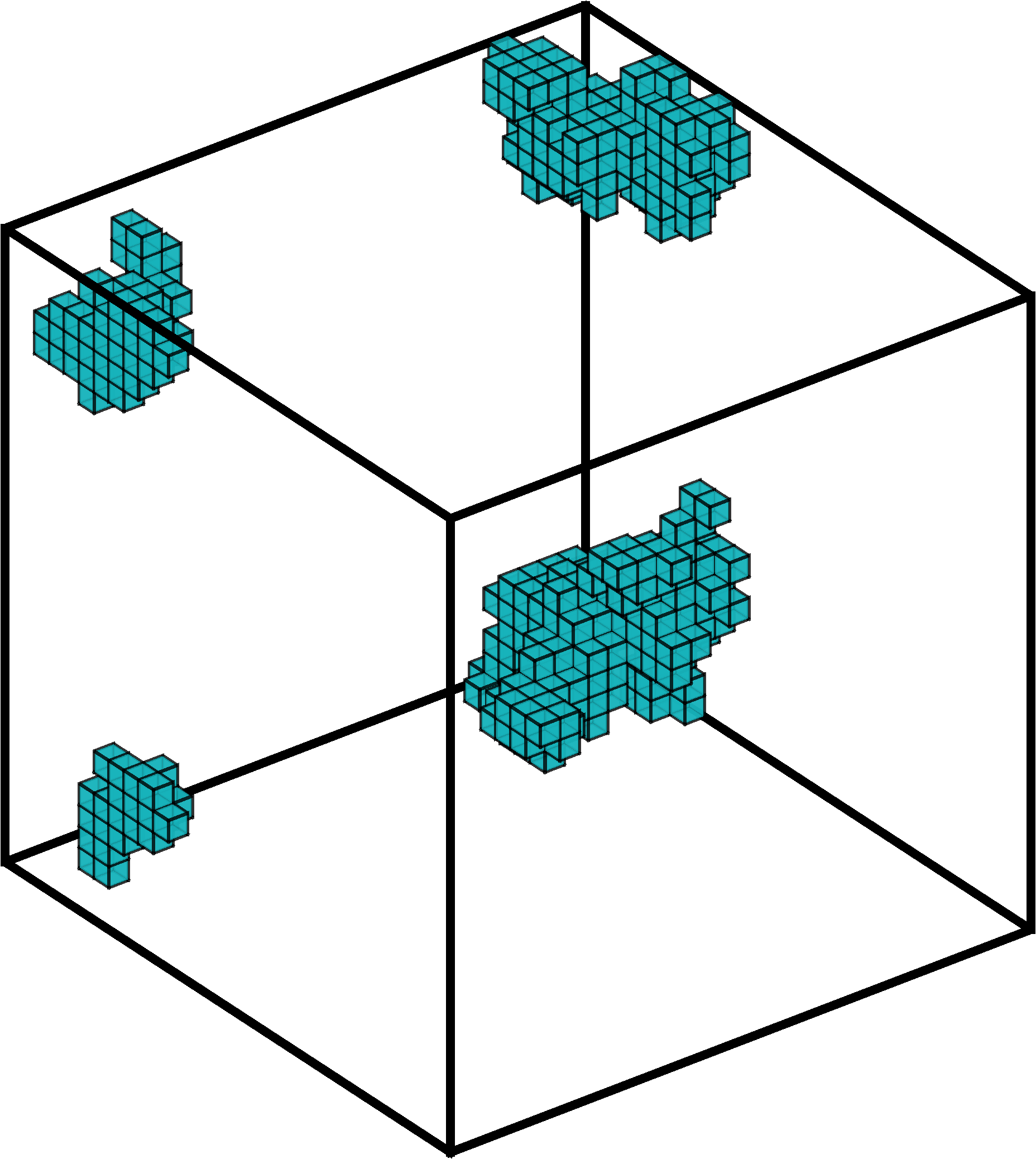}
\includegraphics[width=0.397\columnwidth]{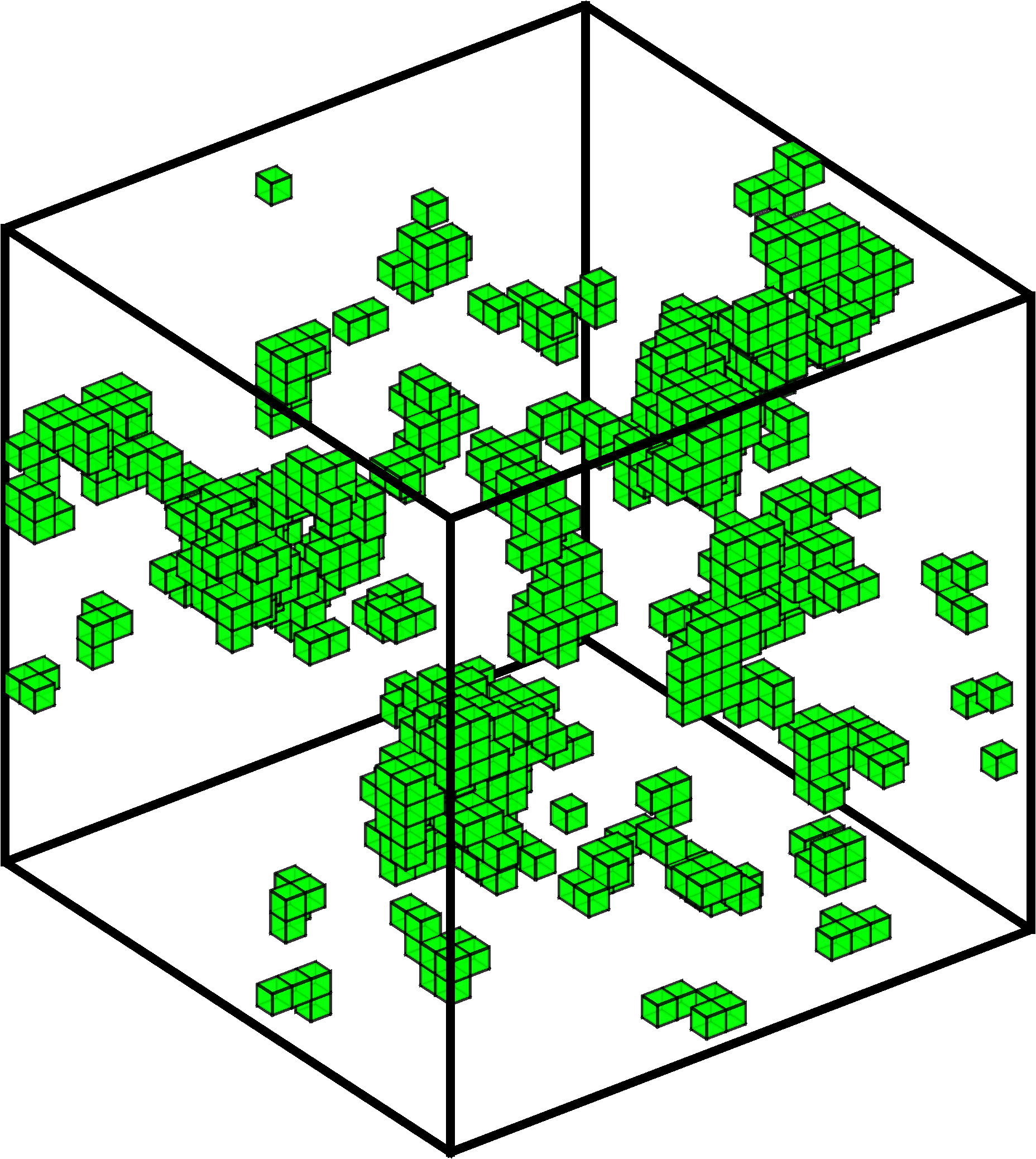}
\includegraphics[width=0.397\columnwidth]{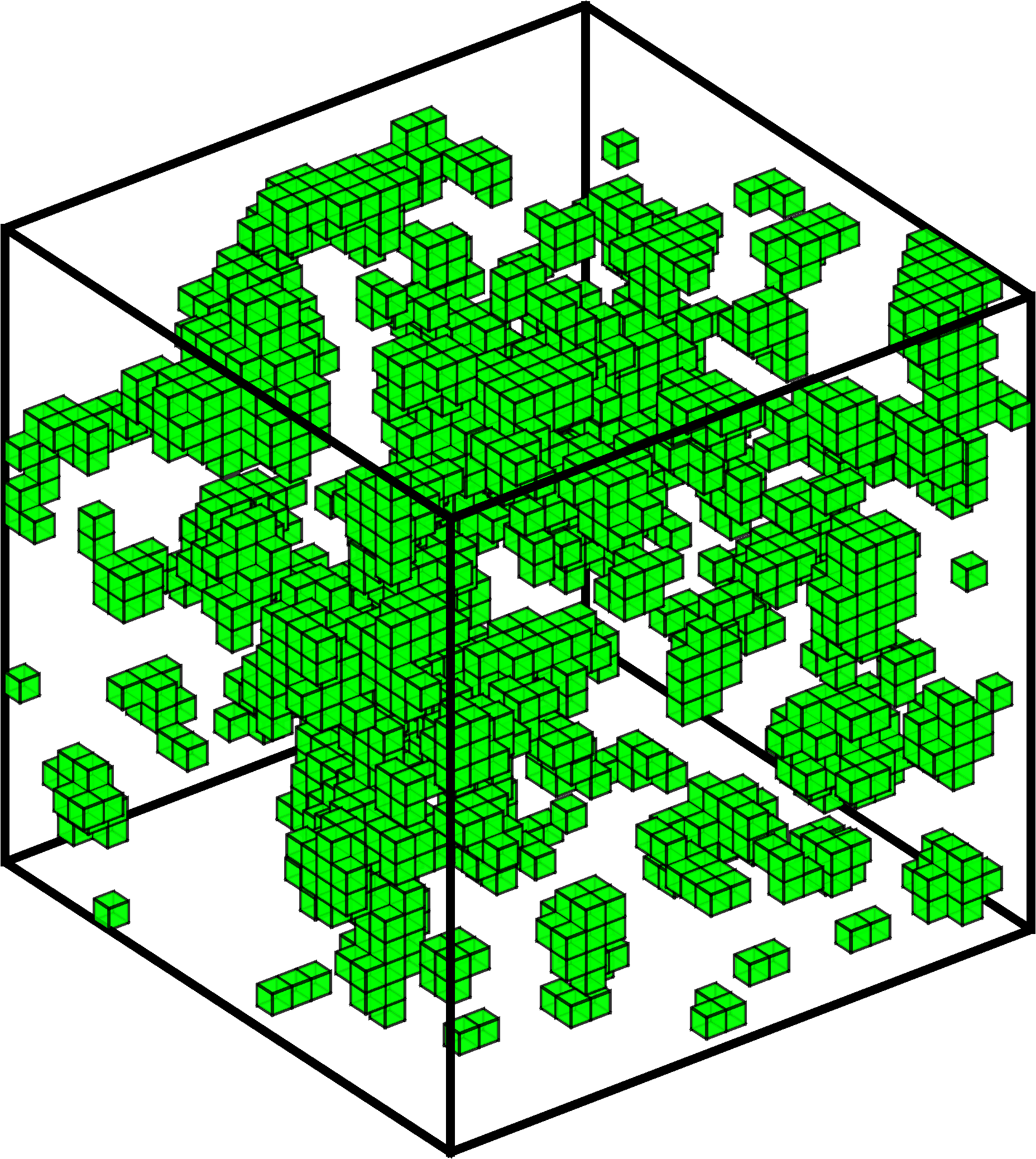}
\caption{\textbf{Jamming vs caging in 3D:} Representative configurations of the 3D model at a) $\rho = 0.37$, which is between the jamming density $\rho_J \approx 0.35$ and the caging density $\rho_C \approx 0.38$, and b) $\rho = 0.41$, which is above $\rho_C$, for a cubic lattice of linear size $L=30$. Frozen particles are marked in dark red, the largest cluster of unfrozen sites in light blue and unfrozen sites that belong to all other, smaller clusters in bright green. For visualization, we separately copied below each image the three components comprising it. At both densities shown here there are frozen particles, however in (a) the largest cluster of unfrozen sites spans the system, while in (b) it is compact. Thus the dynamics are predicted to be diffusive for $\rho<\rho_C$ and caged for $\rho>\rho_C$.}
\label{fig:cluster_map_3d}
\end{figure}

The unfrozen or mobile particles can only travel within connected \emph{clusters} of unfrozen sites. Therefore if there is an infinite cluster of unfrozen sites in the system, we expect to find long-time diffusive behavior, while if all clusters are compact we expect the self-diffusion coefficient to vanish. Hence the caging transition, at which self-diffusion ceases should be related to the percolation of unfrozen sites~\cite{Molinero_CPL_2003,Molinero_PRL_2005,Charbonneau_PNAS_2014,Charbonneau_PRE_2015,Hofling_porous,Kammerer_porous,Schnyder_porous,Spanner2016}. Therefore, for any given random initial configuration of particles on the 3D lattice, we first find all the unfrozen sites, and then identify the clusters of connected unfrozen sites, see Fig.~\ref{fig:cluster_map_3d}. Each mobile particle will be able to travel with time only within the cluster it started in, and thus long-time diffusive behavior requires the existence of an infinite cluster that percolates through the system. We will demonstrate that even when the system is jammed, there may be a percolating cluster of accessible sites and hence the motion of at least a finite fraction of the particles in the system is not restricted. To conclude that their long-time behavior is diffusive and not sub-diffusive we use the fact that for $\rho<\rho_C$ the infinite cluster is not fractal, therefore we expect diffusive behavior in this range of densities, and the self-diffusion coefficient to vanish exactly at $\rho_C$. Precisely at $\rho=\rho_C$ the percolating cluster is fractal and we expect to find subdiffusive behavior~\cite{Sokolov}. 

\begin{figure}[t]
\includegraphics[width=\columnwidth]{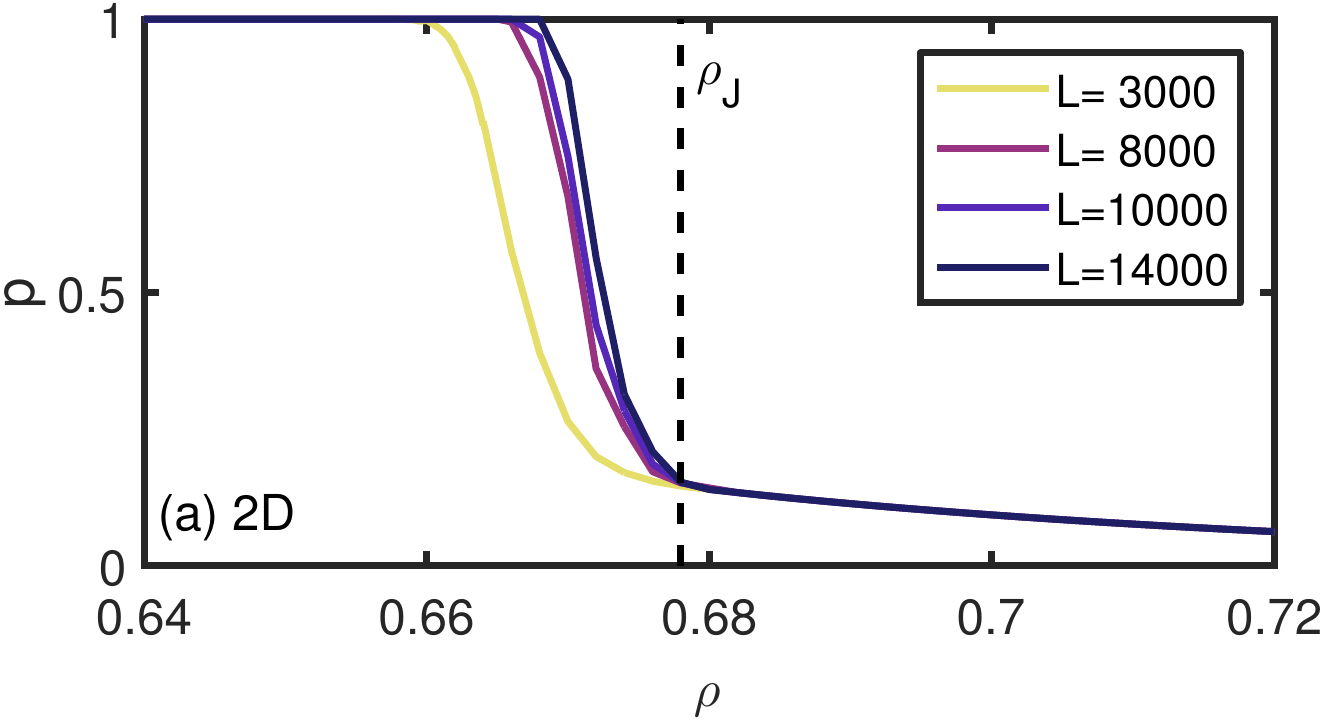}
\includegraphics[width=\columnwidth]{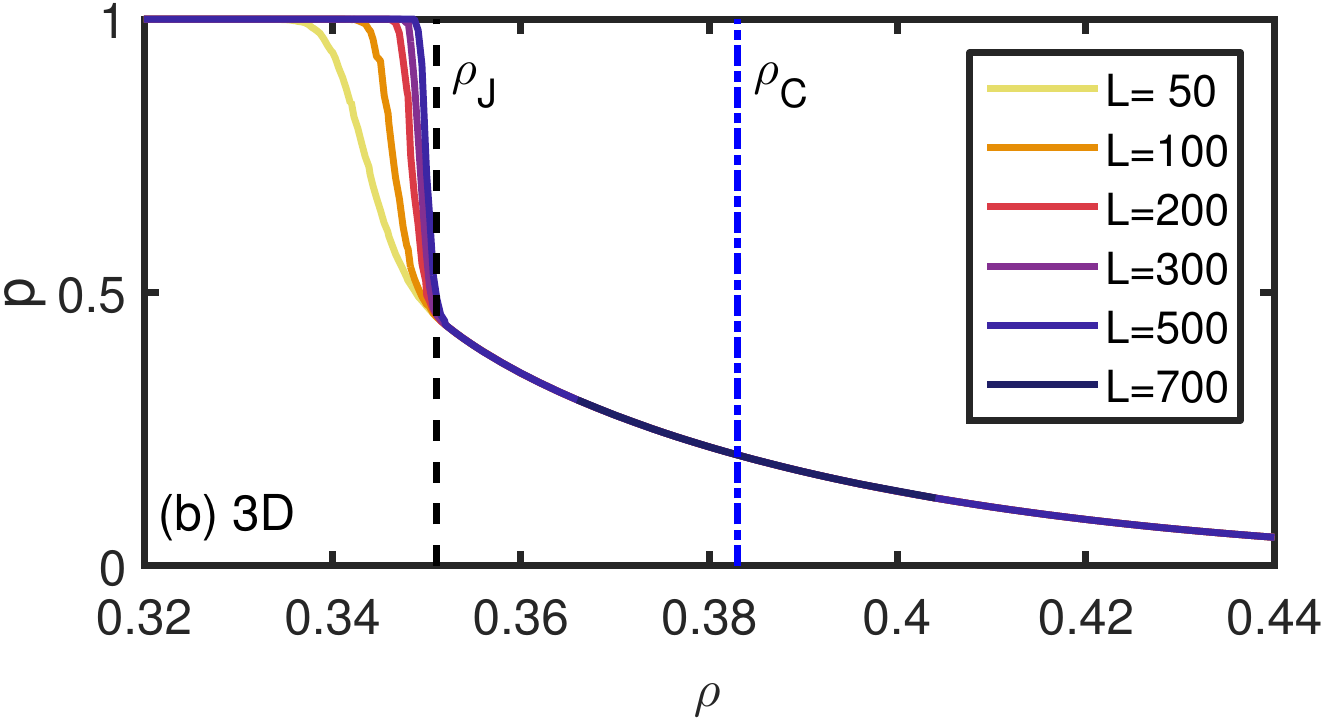}
\caption{\textbf{Accessible sites:} Fraction $p$ of unjammed, or accessible sites decreases with increasing particle density $\rho$. For $\rho<\rho_J$ the system is unjammed and $p=1$. At $\rho_J$ (vertical dashed black line), $p$ jumps discontinuously to a finite value $p_J$, and then decreases smoothly with increasing $\rho$. In 2D (panel a), the system becomes caged at $\rho_J$, while in 3D (panel b), the caging transition (vertical blue dash-dotted line) occurs at some higher density $\rho_C$ and does not exhibit there any singularity in $p(\rho)$.}
\label{fig:p_vs_rho}
\end{figure}

\begin{figure}[t]
\includegraphics[width=\columnwidth]{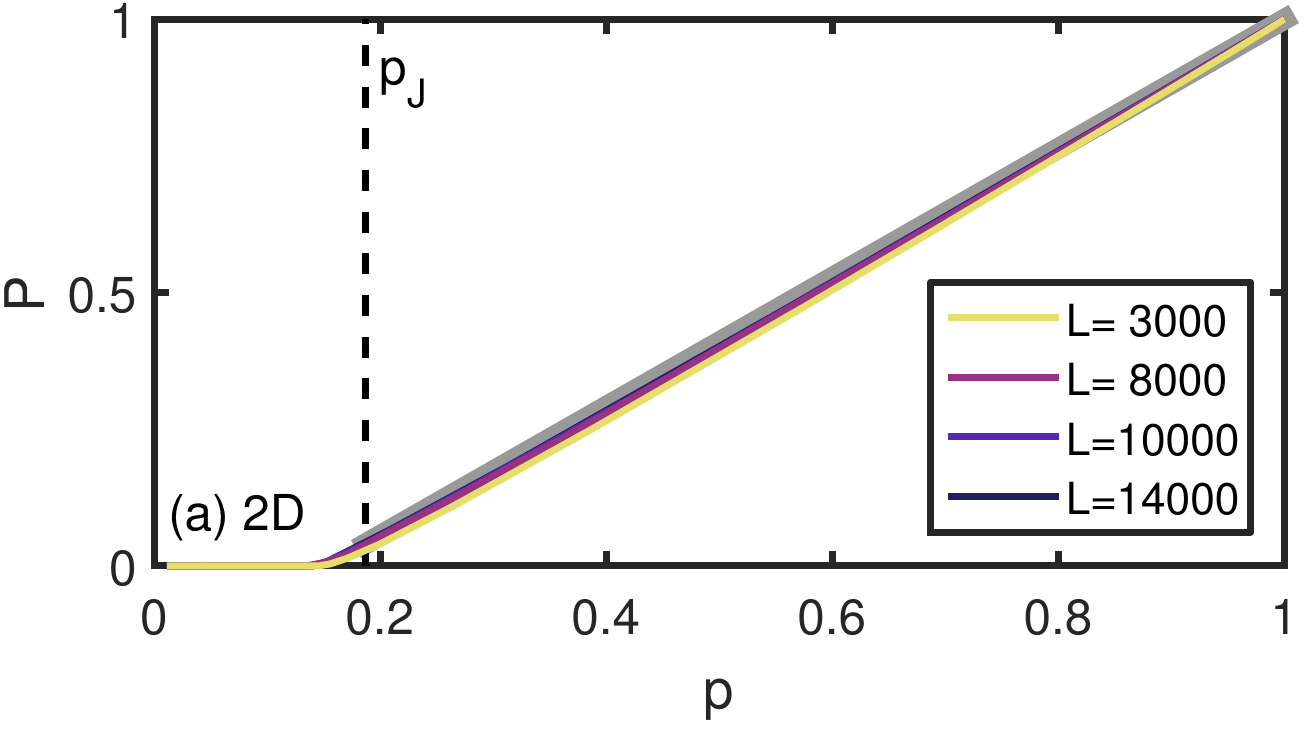}
\includegraphics[width=\columnwidth]{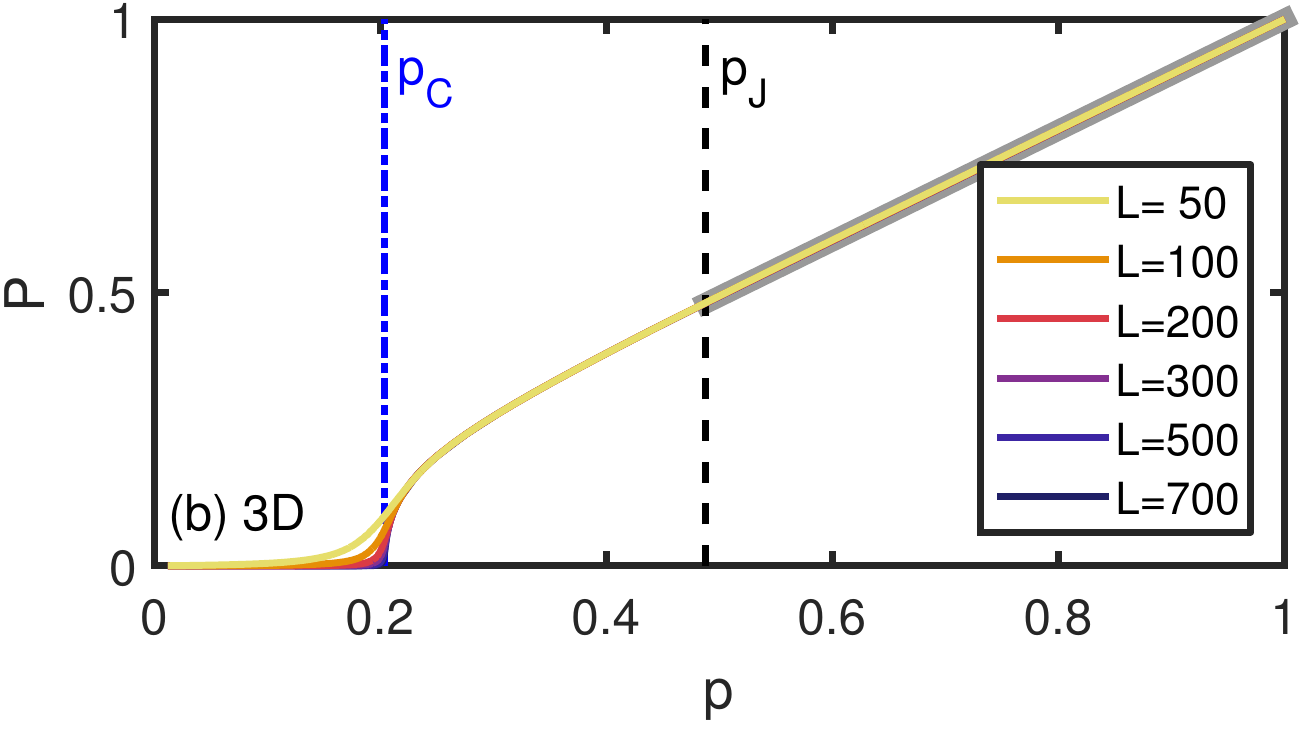}
\includegraphics[width=\columnwidth]{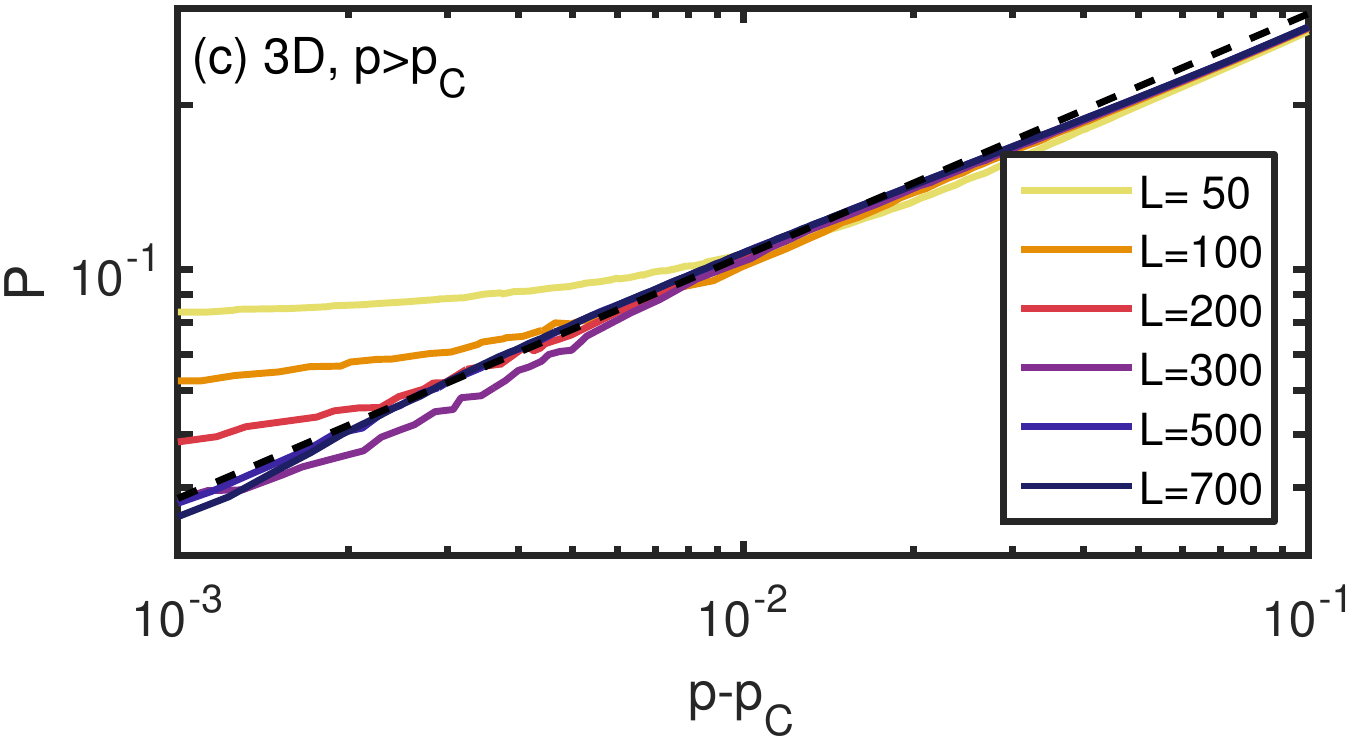}
\caption{\textbf{Infinite cluster:} a,b) Probability that a site belongs to the largest cluster exhibits a continuous percolation transition at $p_C$ (vertical blue dash-dotted line) only in the 3D model. In both 2D and 3D, the behavior for $p>p_J$ is characteristic of a discontinuous transition. Namely, only precisely at $\rho_J$ some of the realizations do not have any frozen particles and some have a finite fraction of frozen particles, thus the measurement averaged over multiple realizations gives a linear interpolation indicated by the thick gray lines. c) In 3D, for $p \gtrsim p_C$, $P$ scales as $P \propto (p-p_C)^\beta$ with $\beta=0.4$ (black dashed line).}
\label{fig:P_vs_p}
\end{figure}

\section{Percolation of unfrozen sites}

In three dimensions, we study the percolation properties of the unfrozen sites as a function of the fraction $p$ of unfrozen (or accessible) sites, which decreases monotonically with the density $\rho$ of particles on the lattice, as shown in Fig.~\ref{fig:p_vs_rho}. For high values of the particle density $\rho$, $p$ is small and the accessible sites are fragmented into many small clusters. For $\rho<\rho_J$ the system is unjammed and thus all sites are accessible, $p=1$ and there is a single infinite cluster covering the entirety of the system. In our 3D model we expect that for some range of densities above $\rho_J$ even though $p<1$, there is an infinite cluster of unfrozen sites, which includes a finite fraction of the sites in the system. Thus we suggest that as $\rho$ decreases and $p$ increases, the clusters of accessible sites become larger and larger until a certain critical value $p_C$ where an infinite cluster forms. For finite systems we use the term infinite for a cluster that includes a number of particles that scales with the system size. To characterize this percolation transition in simulations of finite systems, we begin by measuring the probability $P$ that a site belongs to the largest cluster. Figure~\ref{fig:P_vs_p} shows that in our 3D model $P$ exhibits a continuous percolation transition at $p_c=0.2$ ($\rho_C=0.38$) with $P=0$ for $p<p_C$ and $P$ growing as a power law of $p-p_C$ for $p>p_C$,
\bea
P \propto |p-p_C|^\beta . 
\eea
Not only is this scaling law characteristic of a second-order percolation transition, but the numerical value that we obtain for the exponent $\beta=0.4$ is consistent with the known result $\beta=0.41$ for random percolation~\cite{percolation_book}.

\begin{figure}[t]
\includegraphics[width=\columnwidth]{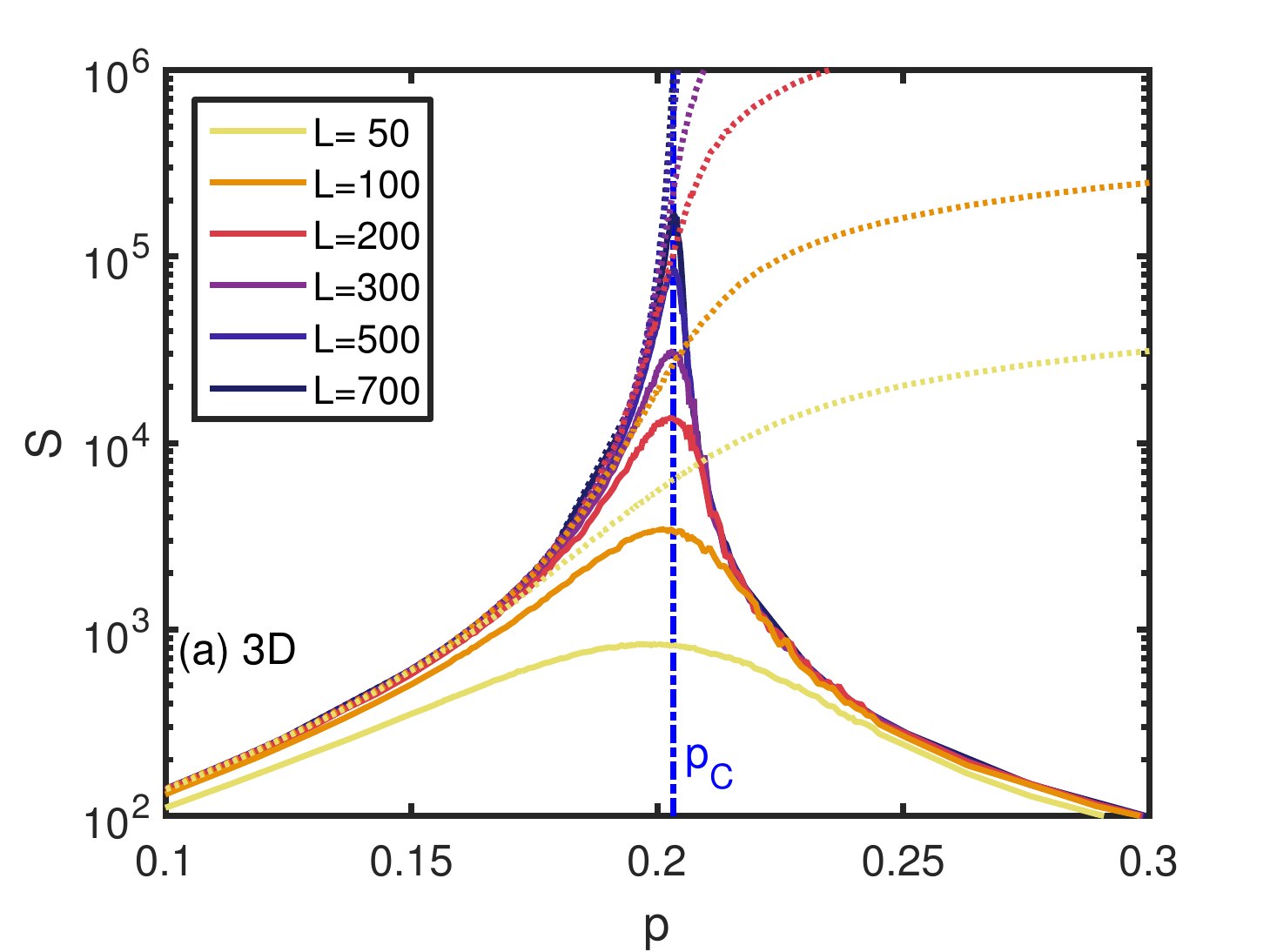}
\includegraphics[width=0.48\columnwidth]{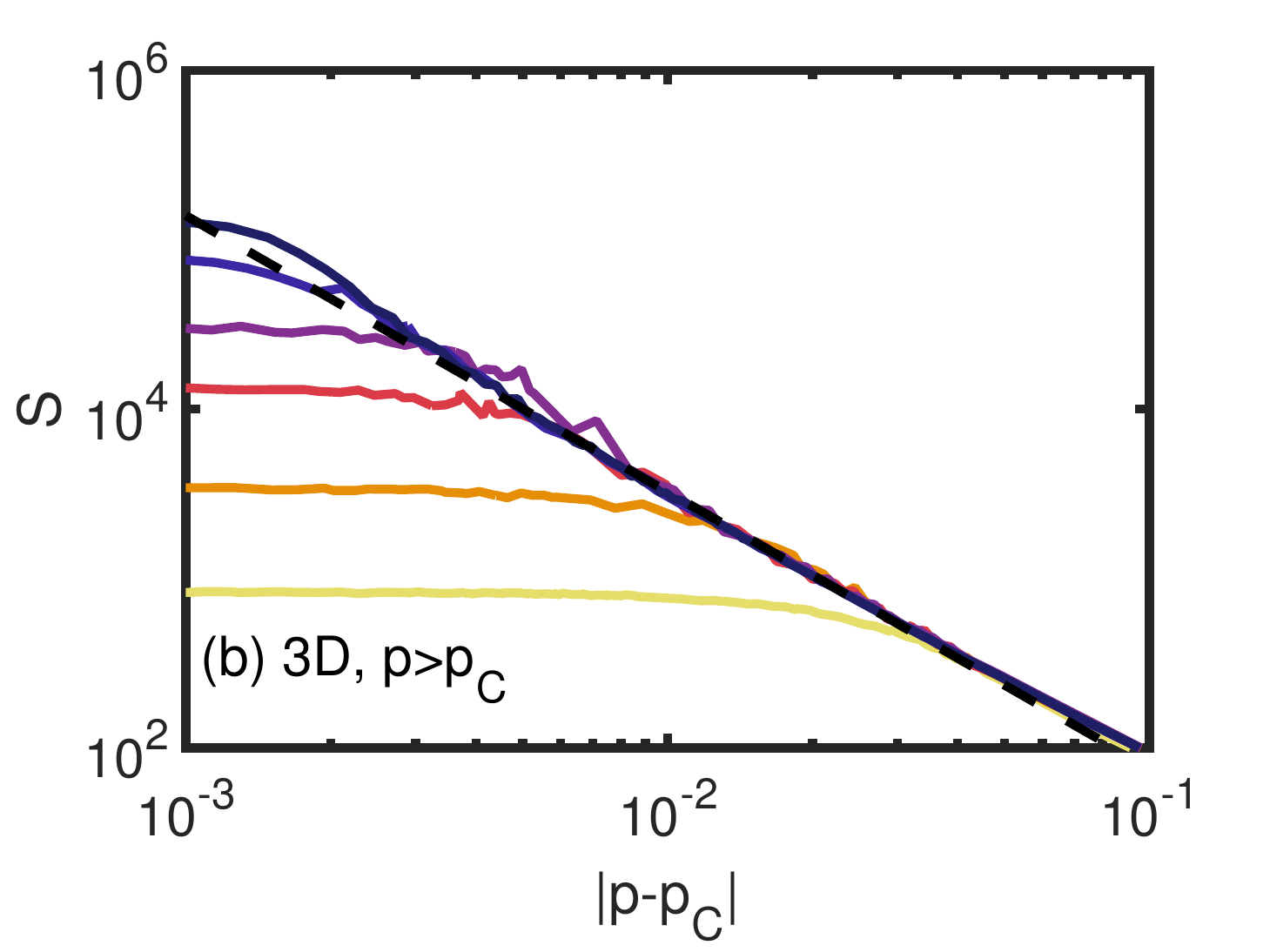}
\includegraphics[width=0.48\columnwidth]{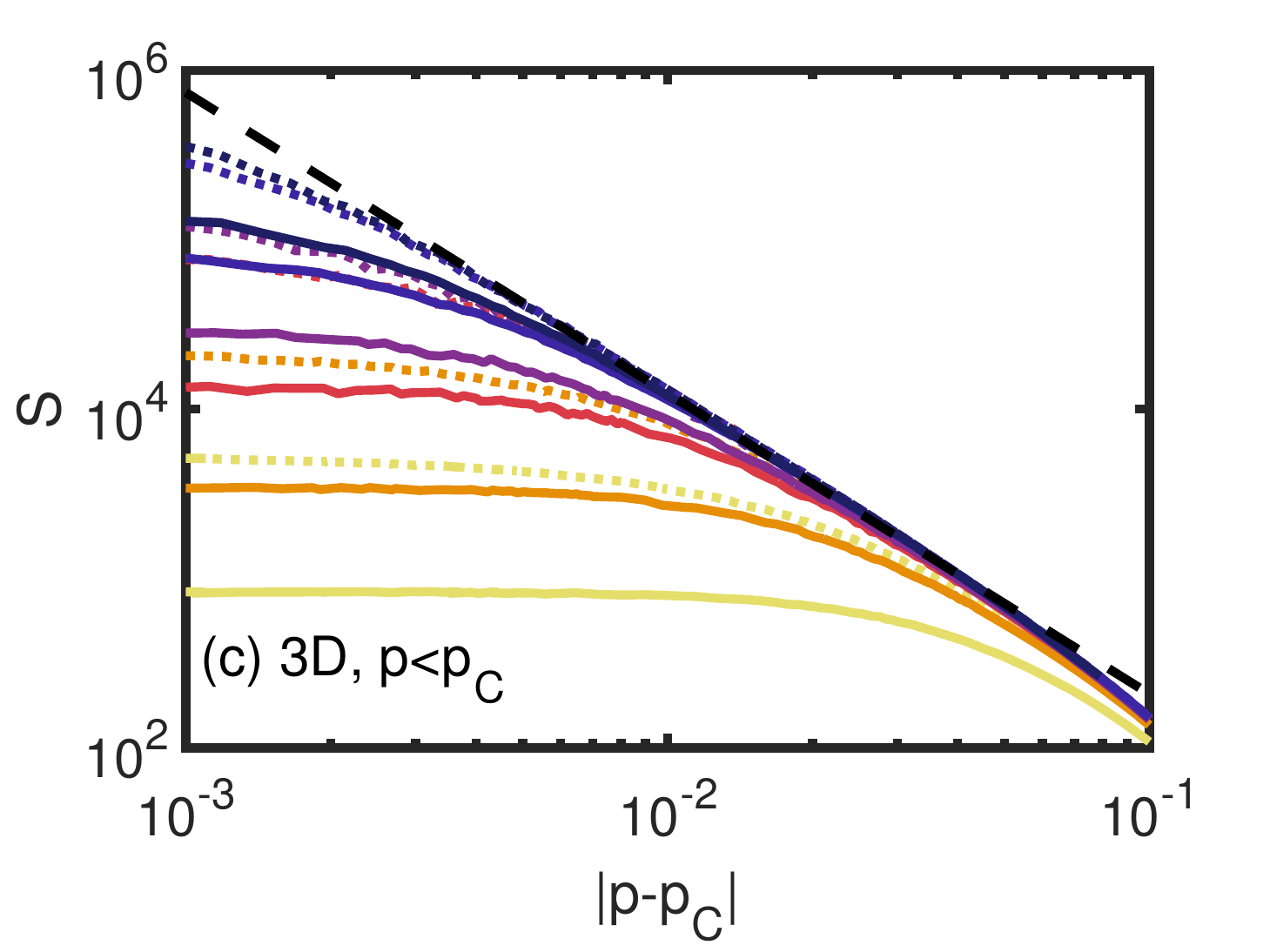}
\caption{\textbf{Finite clusters:} a) The average size of a cluster that each site belongs to in the 3D model. Solid lines are $S_0$ which excludes the largest cluster, and thus peak at the critical density for caging (vertical blue dash-dot line). Dotted lines are $S_1$ which includes the largest cluster, and thus increase with system size for $p>p_C$ since there a percolating cluster exists. b) For $p>p_C$ we find that $S_0 \propto (p-p_C)^{-\gamma}$ with $\gamma=1.6$ (black dashed line). c) For $p<p_C$ we find that $S_0$ (solid lines) and $S_1$ (dotted lines) behave similarly. However since $S_1$ reaches larger values we fit it to  $S_1 \propto (p_C-p)^{-\gamma}$ with $\gamma=1.8$ (black dashed line).}
\label{fig:S_vs_p}
\end{figure}

To identify this critical point at which caging occurs we would like to calculate the average size of the clusters that a site belongs to, excluding the infinite cluster. Since we can numerically consider only finite systems, we define the average sizes $S_0$ and $S_1$ of the cluster that a site belongs to excluding and including the largest cluster, respectively. $S_0$ is the proper quantity for $p>p_C$ and $S_1$ is the proper quantity for $p<p_C$, and in the thermodynamic limit we expect that $S_0 \approx S_1$ for $p<p_C$ and one may use $S_0$ for all the range of $p$.

\begin{figure}[t]
\includegraphics[width=\columnwidth]{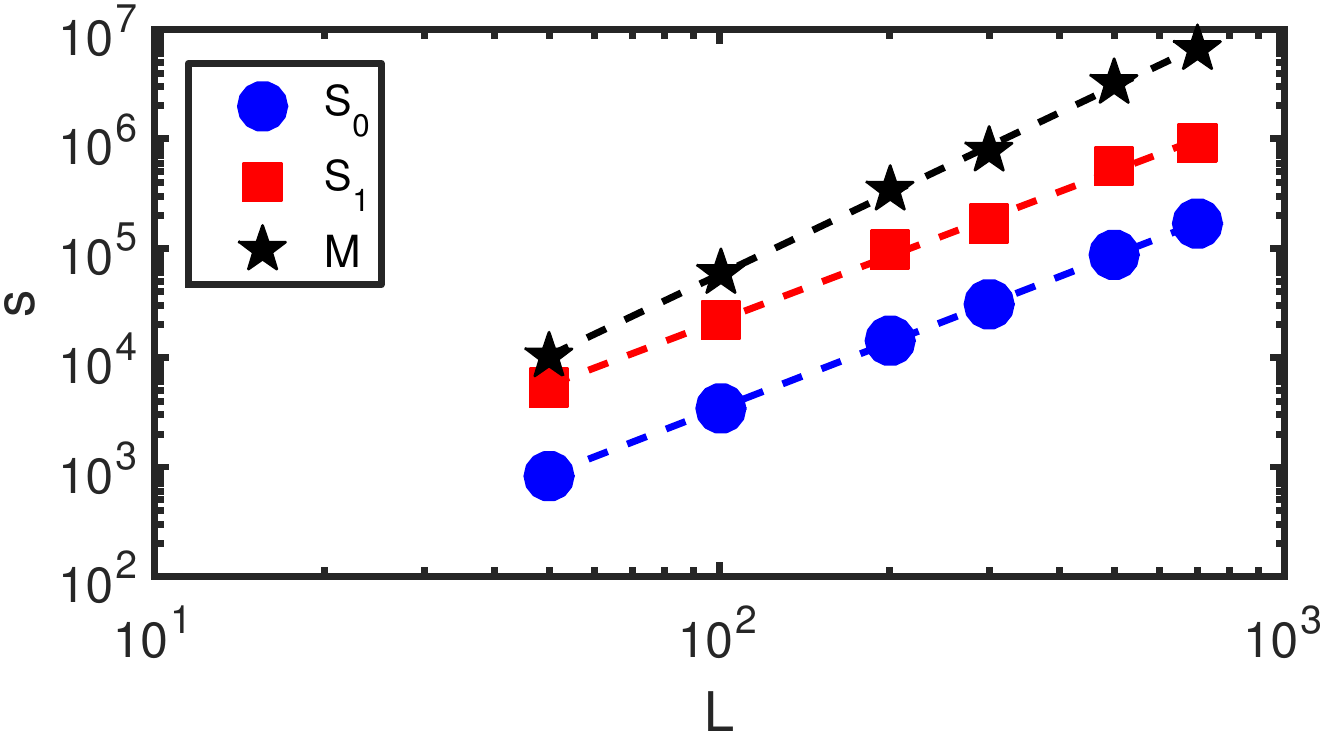}
\caption{\textbf{Maximal cluster:} The average cluster size at the caging transition scales with system size as $S_0(p_C) \propto S_1(p_C) \propto L^2$. The size of the largest cluster scales as $M \propto L^{2.5}$.}
\label{fig:Smax_vs_L}
\end{figure}

Figure~\ref{fig:S_vs_p} shows that as the system size is increased, a sharp peak in $S_0$ develops, and we identify the position of this peak as the caging transition. On both sides of the transition the average cluster size exhibits a power-law scaling with the distance from the critical point. Below the transition ($p<p_C$) we find
\bea
S_1 \propto |p-p_C|^{-\gamma} ,
\eea
with $\gamma=1.8$. Above the transition ($p>p_C$) we find
\bea
S_0 \propto |p-p_C|^{-\gamma} ,
\eea
with $\gamma=1.6$. Our numerical results are not sufficient for determining the critical exponents at a higher accuracy, in particular above the transition where the exclusion of the infinite cluster from the average is not straightforward in finite systems, but we find consistency between our measurement below the transition and the random-percolation value of $\gamma=1.80$~\cite{percolation_book}.

Finally, we test the fractal character of the unfrozen clusters at the critical point $p_C$. In Fig.~\ref{fig:Smax_vs_L} we show that we find that at $p_C$ the largest cluster scales algebraically with the linear dimension $L$ of the system
\bea
M \propto L^{d_f} ,
\eea
with $d_f=2.5$, which is consistent with the random percolation value of $d_f=2.5$. For the average cluster size at the transition we find
\bea
S_0(p_C) \propto S_1(p_C) \propto L^2 ,
\eea
which is consistent with
\bea
S(p_C) \propto L^{\gamma/\nu} ,
\eea
with the random-percolation value of $\gamma/\nu=2.05$ for three dimensions.

\section{conclusions}

In summary, we have numerically studied a 3D jamming-percolation kinetically-constrained lattice-gas model that undergoes two separate phase transitions. At $\rho_J \approx 0.35$ the system jams and a finite fraction of the particles become permanently frozen such that they will never be able to move. Due to the quasi-1D geometry of the frozen structures, we predict that also above $\rho_J$ a finite fraction of the particles in the system will exhibit long-time diffusive motion. These mobile particles travel within a percolating cluster of unfrozen, or accessible sites. We have demonstrated that at $\rho_C \approx 0.38$ this infinite cluster disappears in a continuous phase transition with critical exponents that within our numerical accuracy, are consistent with the random-percolation values. Note however that the value of the critical concentration of unfrozen sites that we find $p_C=0.2$ differs from its value for random percolation on the 3D cubic lattice~\cite{percolation_book}, $p_C=0.312$. The reason for this is in the fact that in our model the configuration of accessible sites is spatially correlated, since their identification is constructed based on the kinetic rules of our dynamical model. 

It is interesting to note the relation to another recently-studied correlated percolation problem, referred to as no-enclaves percolation~\cite{Alvarado2013,Sheinman2015}, in which the structure of clusters requires a geometric support, which has some resemblance to the orientational condition of the kinetic constraint that we study. It would be beneficial to analytically study the caging transition that we have identified and additionally to explore these phenomena in higher dimensions. Moreover, it would be interesting to identify additional models and systems in which jamming is decoupled from caging. A possible direction could be relating our findings to other glassy models exhibiting multiple distinct transitions~\cite{Sellitto2012,Sellitto2013,Sellitto2015,Kondic}.

\section{Acknowledgements}
We thank  Cristina Toninelli, Eli Eisenberg, Oriane Blondel and Yael Roichman for helpful discussions. This research was supported by the Israel Science Foundation grants No. $617/12$, $1730/12$ and by the Prof. A. Pazy Research Foundation.


\begin{thebibliography}{99}
\bibitem{proof_wrong} In the spiral model $\rho_J$ is equal to the critical density of 2D directed percolation~\cite{TBF_Reply,Spiral_EPJB, Spiral_JSP}. In our 3D model we previously claimed that $\rho_J$ is equal to the critical density of 3D directed percolation~\cite{Ghosh}. However, we are aware of a problem in our proof and suspect that the model undergoes a jamming-percolation transition at some lower critical density. A paper clarifying this issue is under preparation.
\end{thebibliography}
\end{document}